\documentclass[pr,twocolumn,showkeys,showpacs,nofootinbib,byrevtex]{revtex4-2}
\usepackage{graphicx,amsmath,amsfonts}
\usepackage{amssymb,amscd,amsxtra,amsthm}
\usepackage{epsfig,bm,dcolumn}
\usepackage{epstopdf}
\usepackage{xcolor}
\usepackage[T1]{fontenc}
\usepackage{lmodern}

\begin{document}

\title{Non-universality of color transparency onset in pion and kaon electroproduction}

\author{Byung-Geel Yu}
\thanks{bgyu@kau.ac.kr}
\affiliation{Center for Exotic Nuclear Studies, Institute for
Basic Science, Daejeon 34126, Korea}
\affiliation{Research
Institute of Basic Science, Korea Aerospace University, Goyang
10540, Korea}

\author{Tae Keun Choi}
\thanks{tkchoi@yonsei.ac.kr}
\affiliation{Department of Physics and Engineering Physics, Yonsei
University, Wonju 26493, Korea}

\author{Kook-Jin Kong}
\thanks{kong@kau.ac.kr}
\affiliation{Research Institute of Basic Science, Korea Aerospace
University, Goyang 10540, Korea}

\begin{abstract}
A combined analysis of the Jefferson Lab data on nuclear
transparency in $A(e,e'\pi^+)$ and $A(e,e'K^+)$ shows that the
onset of color transparency (CT) is not universal across meson
flavors.  The pion transparency is well reproduced by the standard
quantum diffusion model (QDM) with
$\Delta M^2_\pi \simeq 0.7$ GeV$^2$, whereas the kaon data favor
the quadratic expansion of the naive parton model (NPM) with the
natural hadronic scale $R_K \sim \sqrt{\sigma_{KN}/\pi}$.  This
dichotomy cannot be repaired by physically motivated parameter
choices: the pion slope excludes the quadratic expansion with any
physical radius, and the kaon slope requires a QDM excitation scale
far below the range conventionally used in pion transparency
analyses.  A microscopic interpretation, in which the diffusive
evolution of the pion appears as an exceptional consequence of its
Goldstone-boson nature while the kaon follows the generic ballistic
expansion, is discussed.
\end{abstract}
\keywords{Color transparency, Meson electroproduction, Formation dynamics, 
Dynamical chiral symmetry breaking}

\maketitle

Color transparency (CT) is expected to emerge when a
hard exclusive process produces a compact color-singlet
configuration whose interaction with the surrounding nuclear
medium is reduced during its propagation
~\cite{Frankfurt1994,Dutta2013}. Nuclear transparency
measurements in electronuclear reactions therefore provide
a useful phenomenological probe of the space-time
development of the produced hadron
~\cite{Qian2010,Nuruzzaman2011,Das2019,Choi2025,Kong2026}.
While recent measurements of proton transparency have shown
no clear onset of CT up to the highest $Q^2$ measured
~\cite{Bhetuwal2021}, these results further motivate a careful
examination of the meson sector. In this context, the
Jefferson Lab (JLab) data on $A(e,e'\pi^+)$ and $A(e,e'K^+)$
are particularly valuable, because they allow one to compare
the non-strange and strange sectors under broadly similar
kinematical conditions.
This comparison makes it possible to ask not merely whether CT
sets in, but whether its onset is universal across meson
flavors---a question we answer in the negative.

Usually the pion and kaon data are discussed separately.  Their
combined view, however, reveals that the effective formation
dynamics inferred from the observed $Q^2$ dependence differs
qualitatively between the two channels, so that the onset of CT is
not controlled by a universal formation law.  This is the main
message of this Letter; a microscopic interpretation tracing the
difference to the Goldstone-boson nature of the pion is discussed
as an outlook.

Since the two experiments employ deuteron references of different
meaning, we read the published transparencies on a common footing.
Writing the electroproduction cross section on the deuteron as $\sigma^{(\varphi)}_D =
\sigma^{(\varphi)}_p(1+X_\varphi)(1-\delta_D)$ with the Glauber shadowing
correction $\delta_D = (\sigma_{\varphi N}/8\pi)\langle r^{-2}\rangle_D$~\cite{Franco1966}
and $X_\varphi \equiv \sigma^{(\varphi)}_n/\sigma^{(\varphi)}_p$, where
$\langle r^{-2}\rangle_D = 0.31~\mathrm{fm}^{-2}$ is the
inverse-square moment of the deuteron wave function~\cite{Hulthen1957}, the two
normalizations differ:
Ref.~\cite{Nuruzzaman2011} used the deuteron transparency
$T^{(K)}_D = \frac{1+X_K}{2}(1-\delta_D)$ (per
nucleon), whereas Ref.~\cite{Qian2010} used
$T^{(\pi)}_D = (1+X_\pi)(1-\delta_D)$ (single proton).  This reflects
the background filtering---the missing-mass cut removes the
$\gamma^*n\to\pi^+\Delta^-$ channel for $\sigma^{(\pi^+)}_n$, so that $X_{\pi^+} \simeq 0$,
whereas the unresolved hyperon channels leave $X_{K^+} \neq 0$; naive isospin
counting suggests $X_{K^+}=\sigma_{\gamma^*n\to K^+\Sigma^-}/
(\sigma_{\gamma^*p\to K^+\Lambda}+\sigma_{\gamma^*p\to K^+\Sigma^0})\simeq 0.3$--$0.5$.
Because $X_\varphi$ and
$\delta_D$ enter only as a $Q^2$-independent offset,
they cannot
mimic the slope that carries the formation dynamics; we therefore
perform the $\chi^2$ comparison on the hydrogen-normalized $T_A$ and
display $T_{A/D}$ only to confirm that this slope is unchanged under
the deuteron reference.

The central message of the data concerns how the produced compact
configuration is attenuated in the nuclear medium, and is governed
by a single exponent $\tau$.  Created at the hard vertex with a
reduced cross section
$\sigma_h(Q^2) = \sigma_{\varphi N}\,\langle n^2 k_t^2\rangle/Q^2
\sim 1/Q^2$, the compact configuration expands toward its free value
over the formation length $l_f$ as
\begin{equation}
\sigma^{\rm eff}_{\varphi N}(z,Q^2) \simeq \sigma_h(Q^2)
 + \big[\sigma_{\varphi N} - \sigma_h(Q^2)\big]
   \left(\frac{z}{l_f}\right)^{\!\tau}, \qquad z \le l_f,
\label{eq:sigeff}
\end{equation}
reaching the free value $\sigma_{\varphi N}$ for $z > l_f$.  Here $n$
is the number of partons in the compact configuration and
$\langle k_t^2\rangle$ their mean squared transverse momentum, while the
exponent fixes the growth law: $\tau = 1$ is the diffusive
quantum-diffusion evolution, $\tau = 2$ a faster ballistic
expansion.  The onset of CT is thus encoded in a single question---
which value of $\tau$ each channel selects.
In the standard quantum diffusion model
(QDM), the formation length is
\begin{equation}
l_f^{\rm QDM}=\frac{2p_\varphi}{\Delta M_\varphi^2},
\label{eq:lfqdm}
\end{equation}
whereas a simple geometric estimate gives
\begin{equation}
l_f^{\rm NPM}=
\left(\frac{E_\varphi}{m_\varphi}\right)R_\varphi.
\label{eq:lfnpm}
\end{equation}
Here $\Delta M_\varphi^2$ in Eq.~(\ref{eq:lfqdm}) is an
effective squared-mass scale characterizing the coherence and
expansion of the compact configuration, rather than the mass
spacing to a specific physical resonance
\cite{Farrar1988,Larson2006}, while $R_\varphi$ in
Eq.~(\ref{eq:lfnpm}) characterizes the transverse size of the
compact configuration.
For the NPM-based quadratic expansion ($\tau = 2$), the
scale $R_\varphi$ may be associated either with the measured
charge radius or with the characteristic radius implied by
the hadronic cross section $\sigma_{\varphi N}$ \cite{Farrar1988}.
In the comparisons below, we use the latter choice.

\begin{figure}[]
\centering
\hfill
\includegraphics[width=0.44\textwidth]{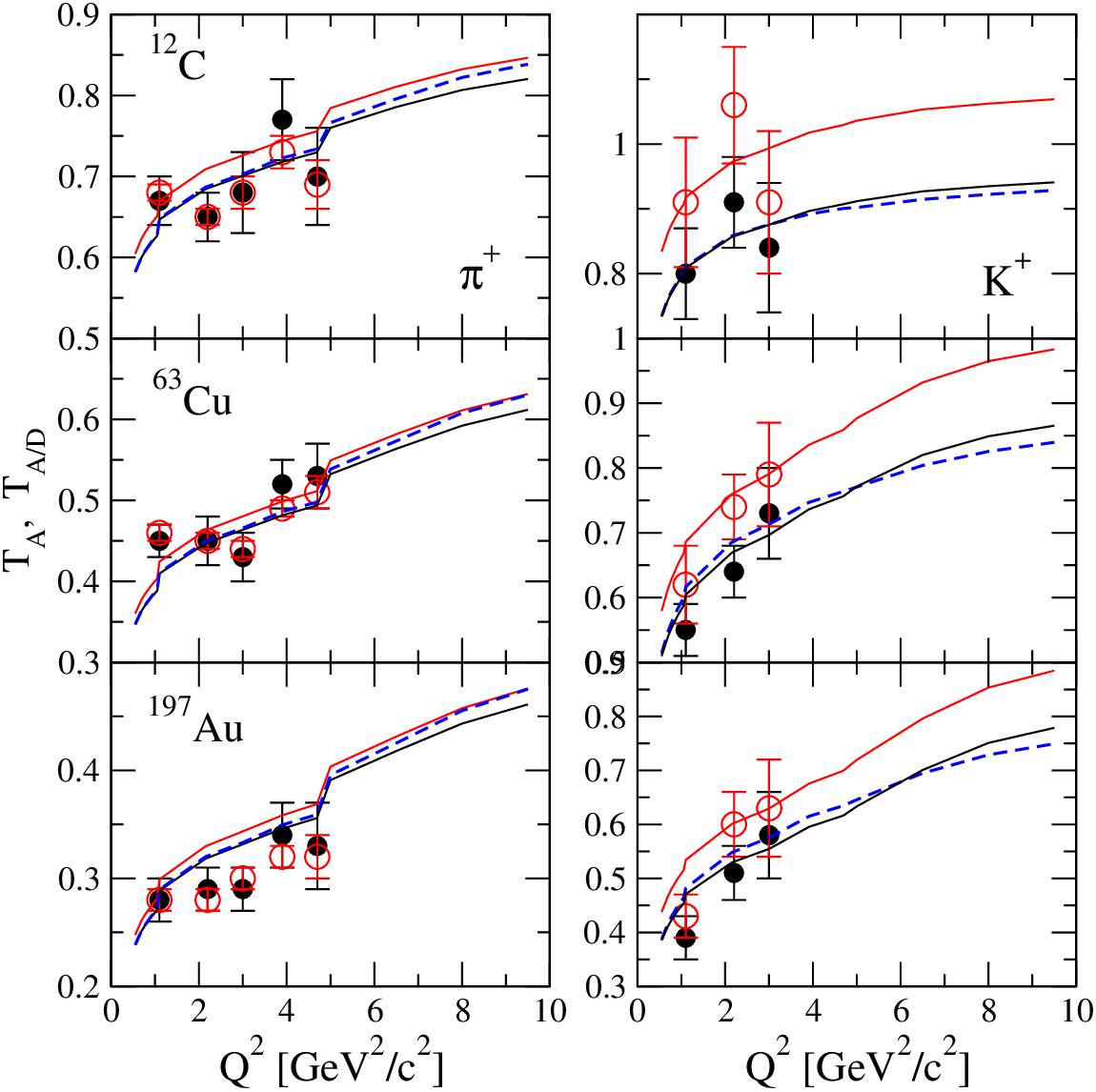}
\caption{ (Color online) Left: nuclear transparency $T_A$ (black)
and deuteron-normalized transparency $T_{A/D}$ (red) for
$A(e,e'\pi^+)$ as a function of $Q^2$ for the indicated nuclei.
The deuteron normalization is evaluated using $X_\pi=0$. The pion
data are well described by the standard QDM with $\Delta
M_\pi^2=0.7~\mathrm{GeV}^2$, whereas the quadratic-expansion NPM
can reproduce a comparable slope only with a very small effective
radius, $R_\pi\approx 0.06~\mathrm{fm}$ (dashed), for the hydrogen
normalization. Right: nuclear transparency $T_A$ (black) and
deuteron-normalized transparency $T_{A/D}$ (red) for $A(e,e'K^+)$
as a function of $Q^2$.
The QDM with the same diffusion scale underestimates the kaon data
~\cite{Nuruzzaman2011,Das2019,Kong2026},
whereas the NPM, implemented through a radius-based quadratic
expansion with $R_K=\sqrt{\sigma_{KN}/\pi}$ and
$\sigma_{KN}=17~\mathrm{mb}$, reproduces the observed rise more
naturally. Within the QDM, a comparable slope is obtained only
with an effective value $\Delta M_K^2=0.15~\mathrm{GeV}^2$
(dashed) for the hydrogen normalization. For the deuteron
normalization, $X_K=0.8$ is used.
Data are from Refs.~\cite{Qian2010} ($\pi^+$) and~\cite{Nuruzzaman2011} ($K^+$).}
\label{fig:piK_CT}
\end{figure}

Figure~\ref{fig:piK_CT} makes the comparison transparent;
the deuteron-normalized curves are evaluated with $X_{\pi^+} = 0$
and $X_{K^+} = 0.8$, the latter
fixed by the deuteron-normalized kaon data, allowing for deviations
from naive isospin counting.
Since $X_{K^+}$ enters only as a $Q^2$-independent offset and the
$\chi^2$ comparison below is carried out on the hydrogen-normalized
data, the value adopted here---though larger than the naive isospin
estimate---does not affect the extracted $Q^2$ slope or any
conclusion drawn from it.
As anticipated, the hydrogen- and deuteron-normalized results differ
mainly by an overall scale, leaving the $Q^2$ slope---the observable
of interest here---almost unaffected.
The pion data are well described by the standard QDM with $\Delta
M_\pi^2 \simeq 0.7~\mathrm{GeV}^2$. By contrast, attempting to
describe the pion data within the quadratic-expansion NPM using
the physical pion size $R_\pi \sim \sqrt{\sigma_{\pi N}/\pi} \sim
0.94~\mathrm{fm}$ derived from $\sigma_{\pi N}\simeq
28~\mathrm{mb}$ \cite{Choi2025} does not reproduce the observed
slope. A comparable trend is obtained only if one assumes an
artificially small radius, $R_\pi \approx 0.06~\mathrm{fm}$,
corresponding to an unphysically small effective cross section. In
this sense, the pion data single out the QDM as the most plausible
description with a physically determined scale.
The kaon channel shows the opposite tendency. The NPM, modeling a
quadratic growth of the compact configuration, works with a
hadronic length scale of natural magnitude, yielding the solid
curve with $R_K \simeq 0.74~\mathrm{fm}$ derived from
$\sigma_{KN}\simeq 17~\mathrm{mb}$ as listed in Table~I of
Ref.~\cite{Kong2026}. By contrast, the standard QDM with the
pion-channel value $\Delta M^2_K \simeq 0.7$~GeV$^2$ falls short
of the observed rise of the data~\cite{Nuruzzaman2011},
as shown in Refs.~\cite{Das2019,Kong2026}.
Forcing the QDM to reproduce the steep kaon slope requires a much
smaller effective mass parameter, $\Delta M_K^2 \simeq
0.15~\mathrm{GeV}^2$ (dashed curve in Fig.~\ref{fig:piK_CT}).
Such a value should be viewed only as an effective parameter
introduced to emulate the observed slope; it is anomalously small
compared with the range conventionally used in QDM analyses of
pion transparency.

In the QDM, $\Delta M^2_\varphi$ should be understood as a
characteristic effective excitation scale associated with the
intermediate configurations through which the compact state evolves
toward its asymptotic size~\cite{Farrar1988,Larson2006}.  It need
not coincide with the squared-mass spacing to any particular radial
resonance; the physical spectrum provides, at most, guidance on the
expected magnitude and flavor ordering of this effective scale.
A complementary light-front estimate provides such guidance without
identifying a specific resonance.  For a $q\bar{q}$ Fock state with
the standard Brodsky--Huang--Lepage wave
function~\cite{Brodsky1983,Huang1994}, the characteristic
light-cone squared-mass gap
$\langle M^2(x,k_\perp)\rangle-m^2_\varphi$, with
\begin{equation}\label{mass}
M^2(x,k_\perp) = \frac{(k_\perp^2+m_q^2)}{x} + \frac{(k_\perp^2+m_{\bar
q}^2)}{(1-x)},
\end{equation}
is \emph{larger} for the kaon than for the pion ($\simeq 1.5$
versus $\simeq 1.2$~GeV$^2$ for constituent masses
$m_{u,d}=0.33$~GeV, $m_s=0.55$~GeV), because the strange-quark
mass raises, rather than lowers, the off-shellness of the
fluctuation.
The light-front estimate places the characteristic kaon scale
above the pion one, giving
$\Delta M^2_K/\Delta M^2_\pi \simeq 1.26$, whereas emulating the
observed slopes within the same QDM parametrization requires
$\Delta M^2_K/\Delta M^2_\pi \simeq 0.2$.
While the absolute light-front values, and the precise ratio, carry
some dependence on the wave-function parametrization, the ordering
$\Delta M^2_K > \Delta M^2_\pi$ is robust: it follows directly from
the strange-quark mass raising the off-shellness in
Eq.~(\ref{mass}), and it is only this ordering and the order of
magnitude, not the absolute normalization, that the argument
requires.
Thus the required
kaon value, $\Delta M^2_K \simeq 0.15$~GeV$^2$, is not only far
below the $\Delta M^2 \simeq 0.7$--$1.4$~GeV$^2$ range commonly
used in pion QDM analyses~\cite{Farrar1988,Larson2006}, but also
opposite in flavor ordering to this simple microscopic estimate.
No established QDM interpretation presently supports such an
exceptionally small kaon excitation scale.
This tension mirrors the pion-channel argument above, where the
quadratic expansion required the unphysical radius
$R_\pi \approx 0.06$~fm.  The pion data therefore strongly favor
the QDM over the NPM when the latter is constrained by a physical
radius, while the kaon data strongly favor the NPM over the QDM
when the latter is constrained by the standard effective excitation
scale.  The non-universality of the CT onset is thus not removed by
physically motivated parameter variations and instead points to a
change in the effective formation law from $\tau=1$ to $\tau=2$.

We have checked that moderate variations of $X_K$ and
$\sigma_{KN}$ within physically reasonable ranges mainly modify
the overall normalization and do not alter the qualitative
conclusion that the kaon data are more effectively described by
the NPM than by the standard QDM.
A $\chi^2$ comparison against the hydrogen-normalized $^{12}$C, $^{63}$Cu, and
$^{197}$Au data quantifies these statements, with all model parameters
fixed a priori.  For the pion (15 points), the QDM with $\Delta
M^2_\pi = 0.7$~GeV$^2$ yields $\chi^2/N = 1.2$, whereas the NPM
with the physical radius $R_\pi = 0.94$~fm gives $\chi^2/N \approx
81$; the artificial radius $R_\pi \approx 0.06$~fm restores
$\chi^2/N = 1.2$, degenerate with the QDM.  For the kaon (9
points), the NPM with $R_K = 0.74$~fm gives $\chi^2/N = 0.9$,
while the QDM with the standard effective scale gives
$\chi^2/N = 1.8$.  While the latter is not, on its own, a decisive
statistical rejection, the only QDM parameter that restores the fit,
$\Delta M^2_K = 0.15$~GeV$^2$ ($\chi^2/N = 1.2$), is the anomalously
small effective excitation scale excluded on the physical grounds
discussed above; the kaon exclusion of the QDM therefore rests on
this physical inadmissibility rather than on the $\chi^2$ difference
alone.
This comparison is not a global refit with channel-specific free
parameters; each scheme is constrained by its standard or
independently determined hadronic scale.

A possible microscopic interpretation of this non-universality is
suggested by the light-cone kinematics already employed
above~\cite{Frankfurt1994,Kopeliovich2002}.  The invariant mass
$M^2(x,k_\perp)$ of the $q\bar{q}$ pair in Eq. (\ref{mass}) plays
a dual role.  Its average over the wave function sets the
diffusion scale $\Delta M^2_\varphi$ discussed above, while
$M^2/(2P^+)$ is at the same time the light-cone Hamiltonian that
drives the spatial evolution of the pair.  The transverse
separation $\Delta r_\perp$ of the quark pair therefore grows with
the corresponding group velocity, $d\,\Delta r_\perp/dz
 = \partial\!\left[M^2/(2P^+)\right]\!/\partial k_\perp$,
so that after a propagation distance $z$
\begin{equation}
\Delta r_\perp \simeq \frac{k_\perp}{x(1-x)\,P^+}\, z,
\label{group-velocity}
\end{equation}
where $k_\perp$ is the transverse momentum, and $x$ and $(1-x)$
are the longitudinal momentum fractions of the quark and
antiquark, respectively.  Note that the quark masses drop out of
the $k_\perp$ gradient of $M^2$, so that the expansion rate is
controlled entirely by the light-cone momentum fractions through
the factor $x(1-x)$; the flavor dependence of the formation
dynamics can therefore enter only through the $x$ distribution of
the wave function, i.e., through the meson PDA.

Equation~(\ref{group-velocity}) makes the generic expectation explicit.  It is a
free-streaming, one-body statement: each light-cone configuration
separates at a fixed velocity, so that $\Delta r_\perp \propto z$
and $\sigma^{\rm eff} \propto \Delta r_\perp^2 \propto z^2$.  The quadratic
expansion ($\tau=2$) of the NPM is therefore not an assumption but
the default behavior of an unimpeded $q\bar{q}$ pair---and no
weighting over configurations can change it, since a weighted sum of
quadratic growths is again quadratic.  Indeed, we have checked that
neither the moments of realistic light-front wave functions, such as
$\langle 1/[x(1-x)]\rangle$ or the dispersion of the expansion rate,
nor the single-channel evolution of the produced wave packet in a
confining light-cone potential~\cite{Kopeliovich2002}  yields an appreciable flavor
distinction or a linear regime; no one-body property of the kaon
singles it out, and the kaon simply follows the generic ballistic
pattern.

The linear law ($\tau=1$) is different in kind: much as an
interference pattern cannot be obtained from any average over
classical trajectories, the diffusive growth $\sigma^{\rm eff} \propto z$
cannot arise from Eq.~(\ref{group-velocity}) at all.  It requires the compact state to
evolve as a coherent superposition of low-lying hadronic states,
\begin{equation}\label{superposition}
|{\rm PLC}(z)\rangle = \sum_n c_n\, e^{-iM_n^2
z/2p_\varphi}\,|n\rangle,
\end{equation}
whose mutual interference---not the motion of any single
configuration---governs the $z$ dependence of
$\sigma^{\rm eff}$~\cite{Frankfurt1994,Farrar1988}.
Truncating Eq.~(\ref{superposition}) to the produced meson
and its lowest excitation
---a single effective state standing for the tower of radial
excitations and continuum, so that $\Delta M^2_\varphi$
is the effective gap of Eq.~(\ref{eq:lfqdm}) rather than
a nominal resonance spacing---
$|{\rm PLC}(z)\rangle \simeq c_0|0\rangle
+ c_1|1\rangle\,e^{-i\Delta M^2_\varphi z/2p_\varphi}$, gives
\begin{equation}
\sigma^{\rm eff}_{\varphi N}(z) = |c_0|^2\sigma_{00}
 + |c_1|^2\sigma_{11}
 + 2|c_0 c_1|\,\sigma_{01}\cos\!\Big(\delta_\varphi
 - \frac{\Delta M^2_\varphi}{2p_\varphi}\,z\Big),
\label{twostate}
\end{equation}
with $\delta_\varphi \equiv \arg(c_1/c_0)$.
Two general consequences of this minimal two-state representation
are worth noting.
First, the small-$z$ expansion of
Eq.~(\ref{twostate}) is linear only through a nonzero relative
phase, with slope $2|c_0 c_1|\,\sigma_{01}\sin\delta_\varphi\,
(\Delta M^2_\varphi/2p_\varphi)$; for real coefficients
($\sin\delta_\varphi=0$) it reduces to the same $O(z^2)$ ballistic
law as Eq.~(\ref{group-velocity}), and the coherent sum adds nothing that a one-body
kinematics could not.
Second, compactness does not forbid this linear term.  For a
representative equal-weight truncation with $\sigma_{00}\simeq
\sigma_{11}\simeq\sigma_{01}\simeq\sigma_{\varphi N}$,
Eq.~(\ref{twostate})
gives $\sigma^{\rm eff}(0)/\sigma_{\varphi
N}=1+\cos\delta_\varphi$, whence
$\sin\delta_\varphi=\sqrt{2\varepsilon-\varepsilon^2}$ with
$\varepsilon\equiv\sigma^{\rm eff}(0)/\sigma_{\varphi N}$ the
compact-configuration cross section at production, i.e.
$\varepsilon \simeq \sigma_h(Q^2)/\sigma_{\varphi N}
= \langle n^2 k_t^2\rangle/Q^2 \simeq 0.1$--$0.3$ over the measured
$Q^2$ range.  Since
$\sin\delta_\varphi$ vanishes only as $\sqrt{2\varepsilon}$---far
more slowly than $\varepsilon$ itself---a strongly compact
configuration still retains an $O(1)$ coherent phase: for these
values the interference already supplies
$45$--$70\%$ of the phenomenological QDM slope $\sigma_{\varphi
N}/l_f^{\rm QDM}$, the remainder coming from higher excitations.
A diffusive ($\tau=1$) onset is thus available whenever the
produced meson couples to its lowest excitation with an
unsuppressed, $O(1)$ coherent phase, and reverts to the ballistic
$\tau=2$ law when that phase is small.  The flavor dependence then
resides entirely in the size of $\sin\delta_\varphi$.
The pion is the natural candidate for such an unsuppressed
$\sin\delta_\pi$.  As the Goldstone boson of dynamical chiral
symmetry breaking (DCSB), it possesses the most strongly dilated
light-front wave function~\cite{Chang2013,Shi2014,Zhang2019} and
the strongest low-energy coupling of the compact configuration to
its lowest excitation among all light hadrons.  A strong, chirally
enhanced coupling does not by itself fix the relative phase; but
it is precisely in the near-chiral regime---where the produced
pseudoscalar and its soft excitations remain dynamically
close---that a real (phase-locked) alignment of the two amplitudes
is least protected, leaving an $O(1)$~$\sin\delta_\pi$ as the
generic expectation rather than a tuned one.  The kaon, displaced
from the chiral limit by the explicit strange-quark mass, has this
chiral enhancement suppressed~\cite{Shi2014}; the coupling
stiffens toward a real alignment, $\sin\delta_K$ shrinks, and the
pair falls back to its default ballistic expansion.
In this sense, the observed non-universality of the CT onset may
be read as a flavor-dependent imprint of DCSB and its explicit
violation on the space-time development of the meson.  Whether this
mechanism quantitatively accounts for the observed phase difference
must be settled by the coupled-channel light-cone Green function
evolution of the produced wave packet~\cite{Kopeliovich2002},
beyond the scope of this Letter.

In summary, the Jefferson Lab pion and kaon transparency data,
placed on a common normalization footing, admit a simple unified
interpretation: the pion follows the standard QDM with
$\Delta M^2_\pi \simeq 0.7$~GeV$^2$, whereas the kaon follows an
NPM-based quadratic expansion with a natural hadronic scale $R_K$.
This channel dependence cannot be absorbed into physically
motivated parameter choices: the kaon slope would demand a QDM
excitation scale far below the range conventionally used in pion
transparency analyses and unsupported by the simple light-front
estimate above, whereas the pion slope would require an
unphysically small expansion radius in the NPM.  The onset of CT in pion and
kaon electroproduction is therefore not universal.  Microscopically,
a generic $q\bar{q}$ pair expands ballistically ($\tau=2$); it is
the pion---whose Goldstone-boson nature sustains the interference
required for diffusive ($\tau=1$) evolution---that is
exceptional, so that the onset of CT carries a flavor-dependent
imprint of dynamical chiral symmetry breaking.

\section*{ACKNOWLEDGMENT}
This work was supported by the Grant No. NRF-2022R1A2B5B01002307
of the National Research Foundation (NRF) of Korea, and by the
Institute for Basic Science (IBS-R031-D1).

\section*{DATA AVAILABILITY}
The data supporting the findings of this study are available within
the article and from the references cited therein.

\end{document}